\documentclass[aps,prb,twocolumn,floats,showpacs,superscriptaddress]{revtex4}
\usepackage{epsfig}
\usepackage{amsmath}
\usepackage{amssymb}

\newcommand{\beq}{\begin{equation}}
\newcommand{\eeq}{\end{equation}}
\newcommand{\bea}{\begin{eqnarray}}
\newcommand{\eea}{\end{eqnarray}}
\newcommand{\nn}{\nonumber}

\newcommand{\eps}{\epsilon}
\newcommand{\veps}{\varepsilon}
\newcommand{\al}{\alpha}
\newcommand{\s}{\sigma}
\newcommand{\lam}{\lambda}

\newcommand{\be}{\beta}

\newcommand{\vp}{\varphi}
\newcommand{\ra}{\rangle}
\newcommand{\la}{\langle}

\newcommand{\ga}{\gamma}

\newcommand{\ua}{\uparrow}
\newcommand{\da}{\downarrow}

\newcommand{\w}{\omega}
\newcommand{\pdag}{{\phantom{\dagger}}}

\newcommand{\wt}{\widetilde}

\begin{document}
\title{Spin-orbital Kondo decoherence by environmental effects
in capacitively coupled quantum dots}
\author{Sabine Andergassen}
\affiliation{Institut NEEL, Centre National de la Recherche Scientifique and
   Universit\'e Joseph Fourier, BP 166, 38042 Grenoble, France}
\author{Pascal Simon}
\affiliation{Laboratoire de Physique et Mod\'elisation des Milieux
   Condens\'es, Centre National de la Recherche Scientifique and
   Universit\'e Joseph Fourier, 38042 Grenoble, France}
\affiliation{Department of Physics and Astronomy, University of Basel,
   Klingelbergstrasse 82, CH-4056, Basel, Switzerland}
\author{Serge Florens}
\affiliation{Institut NEEL, Centre National de la Recherche Scientifique
   and Universit\'e Joseph Fourier, BP 166, 38042 Grenoble, France}
\author{Denis Feinberg}
\affiliation{Institut NEEL, Centre National de la Recherche Scientifique
   and Universit\'e Joseph Fourier, BP 166, 38042 Grenoble, France}

\date{\today}

\begin{abstract}
Strong correlation effects in a capacitively coupled double
quantum-dot setup were previously shown to provide the possibility of both
entangling spin-charge degrees of freedom and 
realizing efficient spin-filtering
operations by static gate-voltage manipulations. Motivated by the use of such a
device for quantum computing, we study the 
influence of electromagnetic noise on
a general spin-orbital Kondo model, and 
investigate the conditions for observing
coherent, unitary transport, crucial to warrant efficient spin manipulations.
We find a rich phase diagram, where low-energy properties sensitively depend on
the impedance of the external environment and geometric parameters
of the system. Relevant energy scales related to the Kondo temperature are also
computed in a renormalization-group treatment, 
allowing to assess the robustness
of the device against environmental effects. 
These are minimized at low bias voltage and for highly symmetric devices, 
concerning the geometry.

\end{abstract}


\maketitle

\section{Introduction}

Controlling and manipulating isolated spins in 
quantum dots has been the subject
of intense research in the last years. One goal of this line of research is the
realization of novel spin-based devices which may 
provide new ways of processing
quantum information.
An architecture for spin-based quantum computing has been proposed by Loss and
DiVincenzo,\cite{loss98} its building blocks being quantum dots made from a
two-dimensional electron gas. Recent developments using time-dependent gates
led to considerable progress, achieving a single spin qubit control
via electron spin resonance in a double-dot (DD) device.\cite{expt-spinqubits}

Another possibility however consists in realizing single or two-spin operations
in a given device, which acts as a logical gate, 
using static control parameters
only, such as gate voltages or constant magnetic fields. Along this direction,
complex operations such as the production of entangled electron pairs,
\cite{entangled_pairs} or spin teleportation \cite{teleportation}
were proposed.
This scheme, avoiding multiple and synchronized time
manipulations, may increase the processing rate and facilitate the integration
into more complex devices. A promising approach for manipulating spins in
semiconductor nanostructures consists in the use 
of a gate-controlled spin-orbit
(Rashba) coupling.\cite{bychkov,nitta02}
Spin precession has been recently revealed in 
metallic rings with spin-orbit
interaction.\cite{precession}
Another way of controlling the spin was proposed by two of the authors in a
system of two capacitively coupled quantum dots 
under a magnetic field operating with
two extra electrons in the charge sector. This setup implements an
entanglement of spin and orbital degrees of freedom by the realization of
an artificial spin-orbit coupling, fully tunable by a gate voltage only.
When such a DD system is driven into the Kondo regime a spin-orbital Kondo
effect occurs, where each spin flip is associated 
with an orbital flip, {\it i.e.} with
an electron transfer from one dot to the other through the leads.
Owing to the unitary transport that takes place below the Kondo temperature,
this property was exploited to propose an 
efficient ({\it i.e.} high-conductance)
Stern-Gerlach spin filter separating an 
unpolarized spin current into two polarized
ones with opposite magnetization.\cite{apl}
Moreover, we demonstrated that such a DD operates 
as a Stern-Gerlach interferometer
(SGI) in presence of a single common lead.
A coherent spin precession can be supplied by
an Aharonov-Bohm (AB) flux obtained by slightly tilting the in-plane magnetic
field,\cite{sg} allowing for a controlled 
realization of a one-qubit phase gate on
spin qubits.

Since the efficiency of the above device is directly related to the
possibility of cooling the system below the Kondo temperature, and hence of
forming a strong-coupling Kondo resonance, the question arises how an external
electromagnetic environmental affects the Kondo physics in the case
of strongly entangled spin-orbit quantum states.
The study of single and double tunnel junctions, as well as of transistors in a
noisy environment modeled by an impedance established more than fifteen
years ago the mechanism of dynamical Coulomb blockade (DCB)
(see Refs.~[\onlinecite{IN,devgrab}] for a 
review). Decoherence effects due to 
electromagnetic environment have been studied 
for series,\cite{aguado} and parallel 
\cite{bruder,dupont}  double dot systems in the 
sequential regime.
The interplay between the Kondo effect and the 
background charge fluctuations has
however not yet been analyzed systematically; the 
issue was addressed only recently
for some specific systems.
Previous investigations of the Kondo effect in presence of DCB focused on
the regime near the charge degeneracy point of a quantum dot coupled to a noisy
back gate.\cite{karyn1,karyn2,BZS,BZGG}
In this case a Kondo model for the charge degree of freedom can be derived,
including a direct coupling of the charge 
variable to the dissipative environmental
modes.
This generates a competition between the Kondo 
screening of the charge doublet by
the electrons and the localization effect due to the ohmic environment.
On the other hand, the physics and the transport 
properties for the conventional Kondo effect
in the spin sector under an electromagnetic noise 
has also been analyzed recently
and appears to be more subtle. The case of an ac
excitation was treated in 
Refs.~[\onlinecite{kaminski00,lopez01}], while 
that of an ohmic noise was
addressed in Ref.~[\onlinecite{florens06}].
This study showed that an ohmic resistance of the environment exceeding half
the quantum value $R_K=h/e^2$ induces a 
suppression of the inter-lead Kondo interactions,
without however preventing the formation of a 
strong-coupling state due to the remaining
intra-lead processes. Transport through the device is therefore suppressed in a
way similar to the usual DCB, while spin 
screening can be completely or partially 
preserved.
For an environmental resistance smaller than 
$R_K/2$ the Kondo effect can normally
develop between the localized spin and both leads.
However, the fully transparent fixed point with 
unitary conductance $G=2e^2/h$ is
only stable when particle-hole symmetry is maintained by the dot plunger gate
voltage.
These striking results imply that, even though 
the Kondo screening of the local spin
survives, non-linear transport properties through
the device appear in general at low-temperature, 
and should be revealed by tuning a
strong environmental impedance.

The present study aims first at merging these 
previously separate analyses of the charge and
spin Kondo effects in the presence of an ohmic 
environment into a detailed study of
background charge fluctuations in a Kondo model where spin and orbital degrees
of freedom are entangled.
The second aspect of this work will highlight how 
the electromagnetic noise may affect
the spin-orbital Kondo effect and, as a 
consequence, reduce the potential performance of
logical spin operations.

The paper is structured as follows. In 
Sec.~\ref{sec:orbital} we introduce the model
and derive the mapping to a general spin-orbital Kondo model characterized by a
maximal entanglement of spin and orbital degrees of freedom.
In Sec.~\ref{sec:circuit} we present a phenomenological description of the
environmental fluctuations using the equivalent 
circuit theory. This description
enables to derive an effective low-energy model that combines both the spin, or
equivalently the orbital fluctuations, and the 
background electromagnetic noise, assumed
to be ohmic in the present analysis. In Sec.~\ref{sec:decoherence} we provide a
perturbative renormalization-group (RG) analysis 
for the low-energy Hamiltonian.
In particular, we discuss the decoherence of the 
spin-orbital Kondo effect induced
by the ohmic bath and elaborate the generic 
low-energy phase diagram, that allows for
an interpretation of the complicated flow of the Kondo couplings.
Strong renormalization effects on the Kondo scale 
may be observed for particular
parameters, restricting the use of the device as 
a high-conductance spin filter.
The particular case of a pure orbital Kondo effect obtained in capacitively
coupled quantum dots is finally analyzed.
We briefly summarize our results in Sec.~\ref{sec:conclusions}.

\section{Spin-orbital Kondo effect in a
double quantum dot}\label{sec:orbital}

\begin{figure}[ht]
   \epsfig{figure=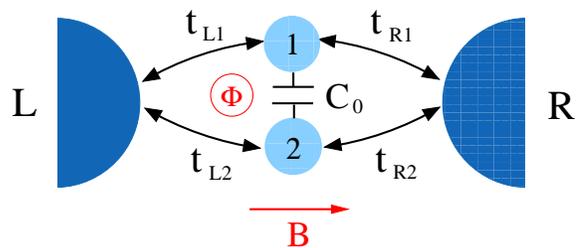,width=7.5cm}
   \caption{(Color online) Schematic 
representation of the proposed setup: two small
     quantum dots coupled by a capacitance $C_0$ and connected to left and right
     reservoirs. Depending on the choice of the 
gate voltages, the upper branch filters
     spin-$\ua$ and the lower one spin-$\da$, or 
vice versa. A magnetic flux $\Phi$
     threads the whole device.}
   \label{Fig:fig1}
\end{figure}

In this section we focus on the orbital Kondo effect in the DD.
We are particularly interested in the spin splitter or SGI introduced
previously.\cite{apl,sg}
The typical device we have in mind is depicted in Fig.~\ref{Fig:fig1}.
Nevertheless, this device is rather generic and encompasses other geometries
based on capacitively coupled quantum dots where orbital Kondo physics is
expected.\cite{borda,pohjola,logan}

\subsection{Derivation of the model}

The number of electrons on a single orbital level 
in each dot is controlled by two
plunger gate voltages via the capacitances $C_{g1}$ and $C_{g2}$, and by a
coupling capacitance $C_0$. The charge states are labeled by $(n_1,n_2)$
extra electrons in dot $1$ and $2$ respectively.
Both dots are connected to the same left and right lead with tunneling hopping
parameters $t_{L 1/2}$ and $t_{R 1/2}$ respectively.
The regime we consider throughout this paper is characterized by the situation
where the low-energy charge states $(1,1)$ and $(0,2)$ are almost
degenerate.\cite{apl} These correspond to eight 
spin-charge states, whose degeneracy
will however be lifted by Zeeman and orbital splitting as discussed below.
For energies lower than the charging energy of the dot
$E_c={\rm min}(E(0,1)-E(1,1);E(1,2)-E(1,1))$ the 
charge dynamics is restricted to these
states only, the states $(0,1)$ and $(1,2)$ occurring as virtual processes.
The lowest excited states $(0,\ua)$ and 
$(\ua,\ua\da)$ are assumed to involve the same
charge excitation energy $E_c=\frac{e^2 
C_0}{4C(C+2C_0)}$, with $C=C_L + C_R + C_g$ for
equal dots and $C_{L/R}$ left/right junction capacitances.

At low energy the isolated DD system is described by \cite{borda}
\beq
\label{hdot} H_{\rm dot}=-\delta E T^z-t T^x-g\mu_B B S^z\;,
\eeq
\noindent where we defined the orbital pseudospin as
\beq
T^z=(n_1-n_2+1)/2=\pm1/2\;.
\eeq
Here $\delta E=E(0,2)-E(1,1)$ is zero when the two lowest charge states are
exactly degenerate. The second term in 
Eq.~(\ref{hdot}) represents a small parasitic
tunneling amplitude between the dots \cite{apl} 
and the last term is the Zeeman splitting
in presence of a local magnetic field applied in 
the $z$ direction. We assume that the
Zeeman energy is large enough so that the total spin-charge states of the setup
amount to the two degenerate ground states $(\ua,\ua)$ and $(0,\ua\da)$.
Note that a large level spacing $\delta\eps$, or equivalently
small quantum dots are necessary to neglect triplet states.\cite{recher,hanson}
When this condition is fulfilled, the total spin $S^z=S^z_1+S^z_2$ is maximally
entangled with the
orbital pseudospin since $S^z=T^z$. As a consequence a spin flip always
involves an orbital pseudo-spin flip.
Note that quantum coherence between spin and 
orbital degrees of freedom has been put forward
in some SU(4) Kondo effect.\cite{borda,lehur03,jarillo,choi05}

Therefore, this setup provides a realization of 
an artificial and fully controlled
local spin-orbit coupling. Besides a stationary 
spin-$\ua$ electron in the lower dot,
the low-energy Kondo screening of the spins by the reservoirs
involves spin-$\ua$ electrons in the upper path 
and spin-$\da$ electrons in the lower one
and vise versa,
corresponding to a maximal entanglement of spin and orbital (or dot) degrees of
freedom. This property has been applied in the 
proposal of an efficient spin splitter in
Ref.~[\onlinecite{apl}] and in the realization of 
single spin operations.\cite{sg}

The full Hamiltonian is
\beq
H=H_{\rm dot}+H_{\rm leads}+H_{\rm tun}\;,
\eeq
where the leads are described by the Hamiltonian
\beq
H_{\rm leads}=\sum\limits_{k,\ga,\s}\veps_k c^\dag_{k,\ga,\s}
c_{k,\ga,\s}^{\phantom{\dag_{k,\ga,\s}}}\;,
\eeq
$ c^\dag_{k,\ga,\s}$ creates an electron with 
energy $\veps_k$ and spin $\sigma$
in the lead $\ga=L,R$ (the Zeeman splitting in 
the reservoirs is negligible).\cite{recher}
The tunneling terms between the leads and the dots are given by
\beq
\label{htun} H_{\rm tun}=\sum_{k,\ga,j,\s} (t_{\ga j}  c^\dag_{k,\ga,\s}
d^\pdag_{j,\s}+{\rm H.c.})\;,
\eeq
where $d^\pdag_{j,\s}$ destroys an electron with 
spin $\s$ in dot $j=1,2$. Note that
the orbital pseudospin is expressed as
$T^z=S^z=(d^\dag_{1,\ua}d^\pdag_{1,\ua}-d^\dag_{2,\da}d^\pdag_{2,\da})/2$.
In order to determine the effective coupling between the DD and the leads,
we consider virtual excitations of the two 
excited states $(1,2)$ and $(0,1)$ due to
tunneling processes between the leads and the 
dots. Using second-order perturbation
theory \cite{sg} in the tunneling amplitudes the Kondo Hamiltonian $H_K$ reads
\begin{widetext}
\bea\label{hk1}
H_{K}=\,\frac{1}{2}\,\left[\phantom{J_{L}^{z\ua}}\!\!\!\!\!\!\!\!\!
\right.&T^z&\left(J_{LL}^{z\ua}\psi^\dag_{L\ua}\psi_{L\ua}^{\phantom{\dag}}-
J_{LL}^{z\da}\psi^\dag_{L\da}\psi_{L\da}^{\phantom{\dag}}+
J_{RR}^{z\ua}\psi^\dag_{R\ua}\psi_{R\ua}^{\phantom{\dag}}-
J_{RR}^{z\da}\psi^\dag_{R\da}\psi_{R\da}^{\phantom{\dag}}
\right)\nn\\
+&T^z&\left(e^{-i\alpha/
2}J_{LR}^{z\ua}\psi^\dag_{L\ua}\psi_{R\ua}^{\phantom{\dag}}-
e^{i\alpha/2}J_{LR}^{z\da}\psi^\dag_{L\da}\psi_{R\da}^{\phantom{\dag}}
+{\rm H.c.}\right)\\
+&T^{-}&\left.\left(J_{LL}^\perp
e^{-i\alpha/2}\psi^\dag_{L\ua}\psi_{L\da}^{\phantom{\dag}}
+J_{RR}^\perp e^{i\alpha/2}\psi^\dag_{R\ua}\psi_{R\da}^{\phantom{\dag}}+
J_{LR}^\perp\psi^\dag_{L\ua}\psi_{R\da}^{\phantom{\dag}}+
J_{RL}^\perp\psi^\dag_{R\ua}\psi_{L\da}^{\phantom{\dag}}\right)+{\rm 
H.c.}\right]\;,\nn
\eea
\end{widetext}
where $\psi_{\gamma\s}=\sum_{\bf k } c_{\bf k 
,\gamma,\s}e^{i\bf k \bf R_{\gamma\s}}$, and 
$\al=2\pi\Phi/\Phi_0$ with
the flux quantum $\Phi_0$ is the AB phase within 
the loop. The phase factor $e^{i\bf k \bf R_{\gamma\s}}$ 
with $\bf R_{\gamma\ua}=\bf 
R_{\gamma1}$ and $\bf R_{\gamma\da}=\bf R_{\gamma2}$ 
takes into account the distance $\bf 
R_{\gamma\ua}-\bf R_{\gamma\da}$ between tunnel 
junctions and distinguishes the different orbital 
states in the leads.
The operators $T^+$ and $T^-$ flip the orbital 
pseudo-spin. We introduced several Kondo couplings
\beq\label{couplings}
J_{\gamma \gamma'}^\perp\approx \frac{t_{\gamma 1}t_{\gamma' 2}}{E_c}
{\rm~~~ and~~~} J_{\gamma \gamma'}^{z\ua/\da}\approx
\frac{t_{\gamma 1/2}t_{\gamma' 1/2}}{E_c}\;,
\eeq
where $\gamma,\ga'=L,R$.

\subsection{Magnetic field effects}

We note that the $J_{\gamma \gamma'}^{z\ua/\da}$ 
Kondo couplings are spin dependent, except for
$t_{\ga/\ga' 1}\simeq t_{\ga/\ga' 2}$. A similar situation appears for a small
quantum dot in the conventional Kondo regime 
connected to spin-polarized leads. The spin
polarization induces a splitting of the Kondo 
resonance which can be compensated by
an external magnetic field.\cite{martinek}
In the present case, due to the entanglement of 
orbital and spin degrees of freedom,
the compensation for such a geometric asymmetry 
is achieved with an orbital field,
{\it i.e.} by fine-tuning the dot gate
voltages $V_{g1}$ and $V_{g2}$. If not stated differently, we therefore assume
$t_{\ga/\ga' 1}\simeq t_{\ga/\ga' 2}$ in the following.

We can rule out the phase in the Hamiltonian (\ref{hk1}) by defining the new
basis $\tilde \psi_{L/R\ua}=e^{\pm i\alpha/4}  \psi_{L/R\ua}$ and
$\tilde \psi_{L/R\da}=e^{\mp i\alpha/4} 
\psi_{L/R\da}$. In this spin-rotated basis
the Kondo Hamiltonian takes the conventional form
\beq
H_K=\sum\limits_{\ga,\ga'}\sum\limits_{\s\s'} 
J_{\ga\ga'}\tilde \psi_{\ga\s}^\dag\vec T\cdot 
{\vec \tau}_{\s\s'}\tilde 
\psi_{\ga'\s'}^{\phantom{\dag}}\;.
\eeq
The disappearance of the AB phase indicates the absence of interference in the
present geometry under specific conditions discussed below.
Indeed, electrons with a spin-$\ua$ travel 
through the upper dot whereas electrons with
spin-$\da$ take the lower one.
In this regime, corresponding to the unitary 
limit, the spin $\vec S$, or equivalently the
pseudo-spin $\vec T$, is completely screened and 
a spin singlet is formed with the left and
right electrodes.

\section{Inclusion of environmental fluctuations}\label{sec:circuit}

A main source of decoherence relies in the 
circuit electromagnetic fluctuations,
coupling to tunneling events to and from each of 
the two dots. As a consequence the
orbital/spin doublet does not fulfill anymore the 
condition of perfect degeneracy and
the quantity $\delta E$ in Eq.~(\ref{hdot}) dynamically fluctuates around zero.
Here we focus on decoherence effects induced by a strong
ohmic environment.

\subsection{Equivalent circuit}
\begin{figure}[b]
   \epsfig{figure=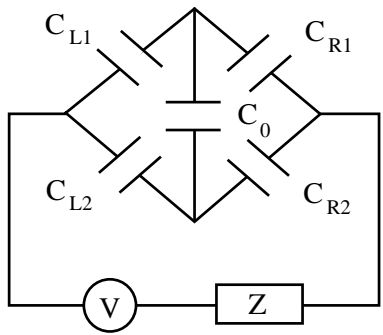,width=5.cm}
   \caption{Schematic representation of the equivalent circuit.}
   \label{Fig:fig2}
\end{figure}

A convenient starting point employs an equivalent circuit representation 
of the SGI
following Refs.~[\onlinecite{IN}] and [\onlinecite{grabert91}].
It contains two single-electron transistors in 
parallel, coupled by a capacitance.
We model the electromagnetic fluctuations by 
introducing an impedance $Z$ closing the
equivalent circuit as depicted in 
Fig.~\ref{Fig:fig2}. We neglect the dot gate
voltage fluctuations as the dot sizes are very 
small and hence $C_{g j}\ll C_{\ga j}$.
Starting from a more complicated equivalent circuit, they can be taken into
account.\cite{grabert91}

Here the environment couples to virtual charge fluctuations as in cotunneling
processes. DCB of inelastic cotunneling through 
two junctions in series was treated
in Ref.~[\onlinecite{odintsov}], while high-order processes
coupling tunneling in two parallel junctions were investigated in
Ref.~[\onlinecite{nazarov91}].

At the Hamiltonian level, electromagnetic fluctuations can be included by the
following transformation of the tunneling amplitudes
\beq
t_{\ga j}\;\to\; t_{\ga j}  e^{i\vp_{\ga j}}
\eeq
in Eq.~(\ref{htun}) and subsequently in Eq.~(\ref{hk1}) through the Kondo
couplings defined by Eqs.~(\ref{couplings}), see Sec.~\ref{sec:eff}.
The phases $\vp_{\ga j}$ are related to the 
voltage fluctuations $\delta V_{\ga j}$
felt by an electron during a tunneling event through the junction
\beq
\vp_{\ga j}(t)=\frac{e}{\hbar}\int\limits_{-\infty}^t \delta V_{\ga
j}(t') dt'\;.
\eeq
The phases $\vp_{\ga j}$ are conjugated of the 
charges $Q_{\ga j}$ on the junction
capacitance $C_{\ga j}$ such that $[\vp_{\ga j},Q_{\ga j}]=ie$ on each dot.
As the phases $\vp_{\ga j}$ originate from the 
same electromagnetic bath, they are
clearly not independent.
Following the general procedure of 
Ref.~[\onlinecite{IN}], the above phases can be
expressed in terms of three phases 
$\vp,\psi_1,\psi_2$, where $\vp$ is the phase
conjugated of the total charge $Q$ seen by the 
impedance, and $\psi_{1/2}$ is the
phase conjugated of the charge in dot $1/2$ respectively.
In the present analysis we consider the experimentally most relevant case of an
ohmic impedance $Z(0)=R$, and define the dimensionless parameter $r=R/R_K$.

We first need to relate the phases $\vp_{\ga j}$ to $\vp$.
The simplest technique involves the equivalent circuit shown in
Fig.~\ref{Fig:fig1}, determining the impedance $Z_{\ga j}$ and the voltage
$V_{\ga j}$ seen by the junction $\ga j$ during a tunneling event.
$Z_{\ga j}$ and $V_{\ga j}$ are obtained in a 
straightforward manner using successive
Thevenin-Norton transformations as described in Ref.~[\onlinecite{IN}].
This yields $Z_{\ga j}=\kappa_{\ga j}^2 Z$ and $V_{\ga j}=\kappa_{\ga j} V$,
where the constant $\kappa_{\ga j}$ depends on the various capacitances via
\begin{widetext}
\beq
\kappa_{L/R,1/2}=\frac{(C_{L,2/1}+C_{R,2/1})C_{R/L,1/2}+C_0(C_{R/ 
L,1}+C_{R/ L,2})}
{(C_{L1}+C_{R1})(C_{L2}+C_{R2})+C_0(C_{L1}+C_{R1}+C_{L2}+C_{R2})}\;.
\eeq
\end{widetext}
We therefore infer that
\beq
\vp_{\ga j}=\kappa_{\ga j}\vp +a_{\ga j}\psi_1+b_{\ga j} \psi_2\;,
\eeq
where the coefficients $a_{\ga j},b_{\ga j}$ are 
determined by the circuit theory.
The phases $\psi_i$ related to the charge on the dots have purely imaginary
correlators at long time and can be discarded.\cite{IN}
Another systematic way to find these results is to compute directly all the
correlators $\la \vp_{\ga j}(t)\vp_{\ga' j'}(0)\ra$ through the involved
transimpedances.
Note that $\kappa_{L,1/2}+\kappa_{R,1/2}=1$.
Indeed, electron transfer processes from one lead to the other without spin
flip affect the phase $\vp$ directly and the system behaves as a double
junction in series.\cite{odintsov,nazarov91}

\subsection{Effective Kondo Hamiltonian}
\label{sec:eff}

The derivation of an effective Kondo Hamiltonian instead involves cotunneling in 
presence of electromagnetic fluctuations as in Ref.~[\onlinecite{florens06}].
We therefore proceed similarly through a 
generalized Schrieffer-Wolff transformation
involving the excitations of the bath degrees of 
freedom. We resort again to the
{\em quasi-elastic} approximation used in 
Ref.~[\onlinecite{florens06}], assuming
that the energy exchanged with the environment 
during the cotunneling process is small
compared to the charging energy $E_c$. Note that 
similar approximations have been
used to treat a single quantum dot in the Kondo regime under an external ac
field.\cite{kaminski00} Under this approximation, 
the Kondo couplings defined in
Eqs.~(\ref{couplings}) are simply dressed by the 
corresponding phase operators and
become
\begin{subequations}
\bea
   J_{\ga\gamma'}^\perp&\to&   \wt 
J_{\ga\gamma'}^\perp =e^{i(\vp_{\ga 
1}-\vp_{\gamma' 2}) }
   J_{\ga\gamma'}^\perp \\
   J_{\ga\gamma'}^{z\ua/\da}&\to & \wt 
J_{\ga\gamma'}^{z\ua/\da}=e^{i(\vp_{\ga 
1/2}-\vp_{\gamma'
       1/2})}J_{\ga\gamma'}^{z\ua/\da}\;.
\eea
\end{subequations}

Therefore, the effective low-energy Hamiltonian 
which encompasses both spin-orbit entanglement 
and environmental
fluctuations has exactly the same form as the 
Kondo Hamiltonian derived in Eq. (\ref{hk1}),
except that the Kondo couplings acquire now a 
dynamical phase, {\it i.e.} $J_i\to \wt J_i$.

We point out the essential property that an originally spin-isotropic Kondo
Hamiltonian becomes generally anisotropic by 
including these charge fluctuations.
This impurity problem is thus more complicated than 
the conventional Kondo problem because of
these dynamical anisotropies, and will be investigated in the following.

\section{Decoherence of the spin-orbital Kondo effect} \label{sec:decoherence}

In this section, we want to study at which degree the spin-orbital Kondo effect
is robust with respect to the electromagnetic background fluctuations.
In the low-energy limit, we can expect at least two phases: i) one where
the spin-orbital Kondo effect fully develops, with a complete screening of the
impurity spin below the Kondo temperature; ii) another one where the spin is
completely localized in one of the $\ua$ or $\da$ states.
In fact, as we will see, and as emphasized previously~\cite{florens06},
transport properties in the Kondo phase also strongly depend on the strength
of the dissipation.

\subsection{Renormalization-group equations}

For symmetric tunneling amplitudes, where the 
indices $1$ and $2$ as well as the
distinction between  $J^\perp$ and $J^z$ becomes superfluous, we recover the
results derived in Ref.~[\onlinecite{florens06}].
This symmetry is not preserved in general for dynamical phases $\vp_{\ga j}$
leading to ten Kondo couplings with $J_{LR}^{z\ua / \da}=J_{RL}^{z\ua / \da}$.
In order to derive the RG flow equations, it may 
be more convenient to introduce
the dimensionless Kondo couplings $\lambda_i=\rho \wt J_i$.
For simplicity we first focus on the case of 
equal initial couplings, {\it i.e.}
$\lam_{\ga \ga'}^{z\ua / \da , \perp}=\lam_0$ for all $\ga,\ga'$;
hence $\lam^{z\ua / \da}_{LL}=\lam^{z\da / \ua}_{RR}$ and
$\lam^{\perp}_{LL}=\lam^{\perp}_{RR}$.
The respective tunneling amplitudes $t_{\gamma,\gamma'}$ are equal for all
$\gamma,\gamma'$ according to Eqs.~(\ref{couplings}), while the capacitances
of the equivalent circuit are assumed as independent parameters.
Nevertheless, the analysis of this particular case allows for an
intuitive understanding of the dynamical emergence
of different phases. Moreover, the corresponding phase diagram turns out
to contain the generic low-energy physics.
The corresponding flow equations read
\begin{widetext}
\begin{subequations}
\label{flow2}
\bea
\frac{d \lambda_{LL}^{z\ua} }{d \ln \Lambda} &=&-
\frac{1}{2}\left[(\lambda_{LL}^{z\ua})^2+(\lambda_{LL}^\perp)^2
+(\lambda_{LR}^{z\ua/\da})^2+(\lambda_{LR}^\perp)^2\right]
\\
\frac{d \lambda_{LL}^{z\da}}{d \ln \Lambda} &=&-
\frac{1}{2}\left[(\lambda_{LL}^{z\da})^2+(\lambda_{LL}^\perp)^2
+(\lambda_{LR}^{z\ua/\da})^2+(\lambda_{RL}^\perp)^2\right]
\\
\frac{d \lambda_{LL}^\perp}{d \ln \Lambda} &=&\beta^2 r
\lambda_{LL}^\perp-\frac{1}{2}\left[\lambda_{LL}^\perp(\lambda_{LL}^{z\ua}+\lambda_{LL}^{z\da})
+\lambda_{LR}^{z\ua/\da}(\lambda_{LR}^\perp+\lambda_{RL}^\perp)\right]
\\
\frac{d \lambda_{LR}^{z\ua/\da}}{d \ln \Lambda} &=&r  \lambda_{LR}^{z\ua/\da}
-\frac{1}{2}\left[\lambda_{LR}^{z\ua/\da}(\lambda_{LL}^{z\ua}+\lambda_{LL}^{z\da})
+\lambda_{LL}^\perp(\lambda_{LR}^\perp +\lambda_{RL}^\perp)\right]
\\
\frac{d \lambda_{LR}^{\perp}}{d \ln \Lambda} 
&=&(1-\beta)^2 r \lambda_{LR}^{\perp}
-\left(\lambda_{LR}^{\perp}\lambda_{LL}^{z\ua}+\lambda_{LL}^{\perp}\lambda_{LR}^{z\ua/\da}\right)
\\
\frac{d \lambda_{RL}^{\perp}}{d \ln \Lambda} 
&=&(1+\beta)^2 r  \lambda_{RL}^{\perp}
-\left(\lambda_{RL}^{\perp}\lambda_{LL}^{z\da}+\lambda_{LL}^{\perp}\lambda_{LR}^{z\ua/\da}\right)\;.
\eea
\end{subequations}
\end{widetext}
These equations explicitely depend on the dimensionless resistance $r=R/R_K$ of
the external circuit, and on the capacitances through the prefactor
$\beta=\kappa_{L2}-\kappa_{L1}$ describing the 
capacitance asymmetry of the system.
The explicit dependence on the capacitances of the equivalent circuit of
Fig.~\ref{Fig:fig2} is given by
\begin{widetext}
\beq
\label{betta}
\beta=\frac{C_{L1}C_{R2}-C_{R1}C_{L2}}
{(C_{L1}+C_{R1})(C_{L2}+C_{R2})+C_0(C_{L1}+C_{R1}+C_{L2}+C_{R2})}\;,
\eeq
\end{widetext}
with $\beta \in [-1,1]$. We focus on $\beta>0$, 
which breaks the symmetry of the system, as a 
sign change corresponds to
an exchange of $L\leftrightarrow R$ lead or $1\leftrightarrow 2$ dot.

The general case for arbitrary initial couplings, taking into account
geometric effects on the capacitances, is discussed subsequently.

\subsection{Symmetric tunneling amplitudes}

\subsubsection{Phase diagram}

\begin{figure}[t]
   \epsfig{figure=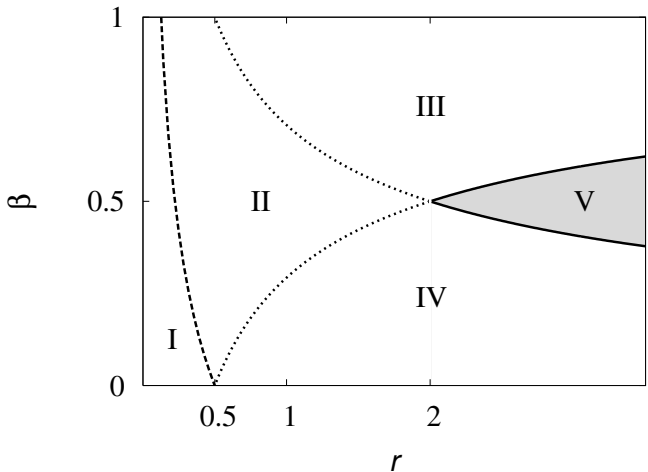,height=8.5cm,angle=-90}
   \caption{Phase diagram as a function of dissipation $r$
     and capacitance asymmetry parameter $\beta$, without 
geometric asymmetry in the tunneling
     amplitudes. The region I corresponds to a robust single-channel
     spin-orbital Kondo effect, and II to a 
complex crossover regime that leads to two
     different Kondo
     dominated regions: III displays a purely orbital Kondo effect, while IV
     is associated with a two-channel Kondo effect 
(for balanced couplings). Finally
     V is a distinct phase where the spin is fully localized (analogous to
     unscreening in the ferromagnetic Kondo 
problem). The solid line indicates a phase
     transition,
     whereas the dashed lines represent a smooth evolution into the crossover
     region II. See text for details.}
   \label{Fig:fig3}
\end{figure}

As detailed below, the analysis of the flow equations~(\ref{flow2})
for various values of $\beta$ and $r$ allows to determine
the phase diagram of the system depicted in Fig.~\ref{Fig:fig3}.

This phase diagram is rather complex, and five different regions within two
distinct phases can be identified.
At small $r$, the region I corresponds to the 
expected spin-orbital Kondo phase, while for
larger $r$, the regions III-IV, characterized by
variations of the Kondo fixed point as discussed subsequently,
emerge from a crossover region II.
The region V corresponds to a different phase and 
arises for both large values of the
dissipation parameter
$r$ and intermediate capacitance asymmetry 
$\beta$; it corresponds to a regime where spin
flips are forbidden by the environment 
(localization). The proximity to this phase is
unfavorable to the spin-orbital Kondo effect, and thus detrimental to the spin
filtering efficiency.

\begin{figure}[hb]
   \epsfig{figure=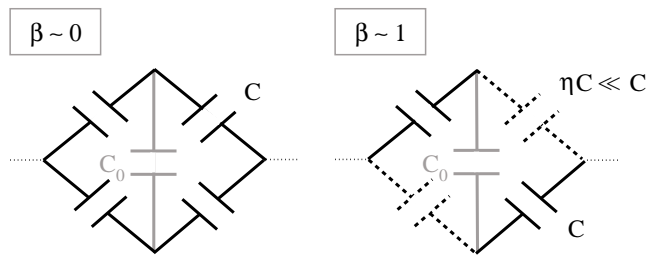,width=8.5cm}
   \caption{Realizations of symmetric capacitance 
configurations for $\beta=0$ and maximal
     capacitance asymmetry for $\beta \lesssim 1$, 
with capacitances $C$ (solid lines)
     and $\eta \,C$ (dashed lines).
     Note that $\beta=0$ includes any configuration of the capacitances
     symmetric under the exchange of the dot 
and/or lead indices independently of
     $C_0$, whereas $\beta \lesssim 1$ is obtained 
for $\eta \ll 1$ and $C_0 \ll C$.}
   \label{Fig:fig4}
\end{figure}

Before studying the generic case for arbitrary $\beta$ we first discuss
the limit $\beta = 0$ corresponding to symmetric 
capacitance configurations. We further
extend the analysis to finite $0<\beta<1$, and finally address the limit of
maximal capacitance asymmetry for $\beta \lesssim 
1$. Realizations of the limiting
cases for a simplified parametrization are shown in Fig.~\ref{Fig:fig4}.

\subsubsection{Symmetric capacitances: $\beta=0$}
\label{sec:sym}

For $\beta=0$ the system~(\ref{flow2}) reduces to two equations describing the
weak-coupling behavior of $\lambda_{LL}$ and $\lambda_{LR}$
\begin{subequations}
   \label{JJ}
   \begin{eqnarray}
     \label{JF}
     \frac{d \lambda_{LL}}{d\log \Lambda} & = & 
-\lambda_{LL}^2 - \lambda_{LR}^2\\
     \label{JB}
     \frac{d \lambda_{LR}}{d\log \Lambda} & = 
&r\lambda_{LR} -2\lambda_{LR}\lambda_{LL}\;,
   \end{eqnarray}
\end{subequations}
as examined in Ref.~[\onlinecite{florens06}].
The inter-lead or "backscattering" Kondo coupling $\lambda_{LR}$ involves
environmental effects via the dynamical phase 
$\vp(t)$, while the intra-lead coupling
$\lambda_{LL}$ instead is not dressed by phase fluctuations within the
quasi-elastic approximation.

\begin{figure}[t]
   \epsfig{figure=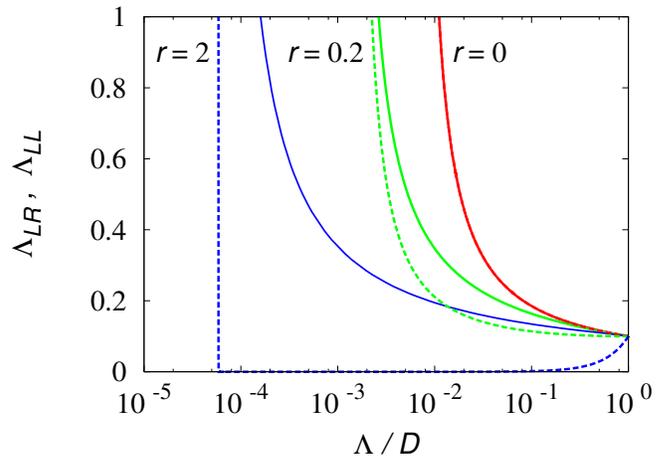,height=8.5cm,angle=-90}
   \caption{(Color online) Flow of the Kondo 
couplings $\lambda_{LL}$ (full line)
     and $\lambda_{LR}$ (dashed line) with the 
running cutoff $\Lambda$, according
     to Eqs.~(\ref{JJ}),
     for different values of $r$ in the symmetric 
case $\beta=0$. The initial values of the
     Kondo couplings at the initial scale $\Lambda_0=D$ (which corresponds to
     the electronic bandwidth) are taken here as 
$\lambda_{LL}=\lambda_{LR}=0.1$.}
   \label{Fig:fig5}
\end{figure}

The results for the flow equations~(\ref{JJ}) for 
different dissipation strengths $r$
are shown in Fig.~\ref{Fig:fig5}.

With increasing $r$ a systematic decrease of the Kondo temperature $T_K$
is observed, marking a delay for the onset of the strong-coupling regime.
Moreover, the appearance of different strong-coupling fixed points can be
identified.
For a small environmental resistance compared to 
the quantum value $R_K$ a non-ohmic
behavior characterized by DCB prevails. A more 
detailed analysis shows that this
result extends also to unbalanced initial couplings.\cite{florens06}
For values of $R$ comparable to $R_K$ the dissipative term in Eq.~(\ref{JB})
controls the RG flow, inducing a suppression of the
inter-lead coupling $\lambda_{LR}$ and therefore 
of the coherent charge transfer
between left and right lead.
Due to the remaining intra-lead processes the coupling $\lambda_{LL}$
(and $\lambda_{RR}$ as well) still renormalize to strong coupling
according to Eq.~(\ref{JF}). This corresponds to a phase where the spin flip
processes are still coherent but charge transfer between the left and right
leads is incoherent.
More challenging from the experimental point of view, a large dissipation
drives the electron tunneling to zero and a non-ohmic regime develops,
characterized in the spin sector by a Kondo 
effect in the most strongly coupled electrode.
For the symmetric case with $\lambda_{LL}=\lambda_{RR}$ a two-channel Kondo
fixed point is stabilized by DCB.
These results are confirmed by a strong-coupling analysis from the
low-energy equivalence of a single Kondo quantum dot in an ohmic environment
described by Eqs.~(\ref{JJ}), and the problem of a $S=1/2$ magnetic impurity
coupled to Luttinger liquid leads.
Two distinctive regimes arise depending on 
$r>r_c$ or $r<r_c$, with the critical
value $r_c=1/2$. These two phases, associated with single and two-channel Kondo
respectively, correspond in the phase diagram depicted in
Fig.~\ref{Fig:fig3} to the phases I and IV.
In particular, phase IV is characterized by anomalous low-temperature transport
properties, accessible in a strong-coupling analysis.\cite{florens06}
Let us now analyze the general asymmetric case $\be\ne 0$.

\subsubsection{Asymmetric capacitances: $\beta>0$}
\label{sec:asym}

\begin{figure}[t]
   \epsfig{figure=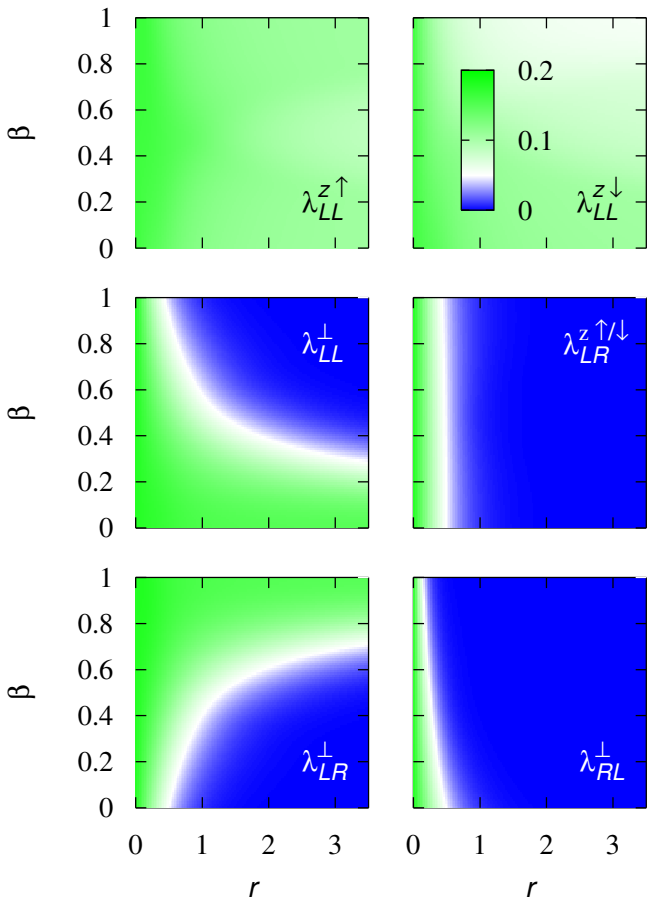,width=8.5cm}
   \caption{(Color online) Typical value of the renormalized Kondo couplings
     $\lambda_i$ according to Eqs.~(\ref{flow2}) 
with equal initial couplings. These
     are displayed as a function of dissipation 
$r$ and asymmetry parameter $\beta$.
     See text for details.}
   \label{Fig:fig6}
\end{figure}

We numerically solved the flow 
equations~(\ref{flow2}) for various values of 
$\beta$ and $r$.
Fig.~\ref{Fig:fig6} shows results for several Kondo couplings as a function
of dissipation $r$ and asymmetry parameter $\beta$. The value for all
the initial Kondo couplings at scale 
$\Lambda_0=D$ was chosen to be $\lambda_i=0.1$, 
and the
presented data correspond to a set of running couplings $ \lambda_i$
at the intermediate scale $\Lambda^*=\sqrt{D 
T_K}$  for better readability, where $T_K$ is
defined by the strongest divergent coupling, at 
the value $\lambda_{LL}^{z\ua}(\Lambda\!=\!T_K)=10$.
This three-dimensional representation of the 
running coupling constants allows us to 
immediately
read which of them are driven to strong coupling 
or are irrelevant in the low-energy limit.

From the different panels we can directly identify several distinct regimes:
$\beta\gtrsim 0$, $\beta \lesssim 1$ and $\beta \simeq 1/2$.

\underline{\em Case $\beta\gtrsim 0$.}\\
For small values of $\beta$ the physical picture is expected to be controlled
by the proximity to the $\beta=0$ line, discussed in Sec.~\ref{sec:sym}.
In this case, the first three equations of the system~(\ref{flow2}),
corresponding to the intra-lead processes, reduce to Eq.~(\ref{JF}), and the
other inter-lead couplings follow Eq.~(\ref{JB}). The dissipation affects more
strongly the latter ones and drives them irrelevant for large values of $r$, as
shown on the fourth to sixth lower panels of Fig.~\ref{Fig:fig6}.
Nevertheless, one can see that these are not driven irrelevant simultaneously
for given values of $\beta$ and $r$. Suppose, we fix for example
$\beta\approx 0.3$, we see that for moderate 
values of $r\simeq 0.8$, $\lambda_{RL}^\perp$
and  $\lambda_{LR}^{z\ua/\da}$ are driven 
irrelevant whereas $\lambda_{LR}^\perp$ is still
going to strong coupling. This defines an 
intermediate crossover regime (indicated as II
in Fig.~\ref{Fig:fig3}) between the single-channel spin-orbital Kondo
phase I, where all three inter-lead processes are 
enhanced at low energy, and the two-channel
phase IV, where these are all irrelevant.

The most pronounced suppression occurs for 
$\lambda_{RL}^{\perp}$, as seen on the
sixth panel of Fig.~\ref{Fig:fig6}. Indeed, in 
the associated flow equation (\ref{flow2}f), a
prefactor $(1+\beta)^2$ dresses the dissipative 
(irrelevant) contribution $r\lambda_{RL}^{\perp}$.
As a rough estimate, the coupling is completely 
suppressed when the total coefficient of this
irrelevant term reaches $(1+\beta)^2 r \simeq 1/2$. This defines the crossover
line between the regions I and II, as reported in the phase diagram on
Fig.~\ref{Fig:fig3}.

Moreover, all the inter-lead couplings are 
completely irrelevant when $\lambda_{LR}^\perp$ 
is also
driven to zero. This occurs for larger $r$ values, as seen in the fifth
panel of Fig.~\ref{Fig:fig6}, and corresponds to the condition
$(1-\beta)^2r\simeq1/2$ from the associated flow equation~(\ref{flow2}e).
This line separates the intermediate crossover 
region II from the two-channel Kondo
phase IV in Fig.~\ref{Fig:fig3}.

\underline{\em Case $\beta\lesssim 1$.}\\
We next consider the strongly asymmetric case $\beta\lesssim 1$.
From the equivalent circuit representation a 
large value of $\beta \lesssim 1$ implies
$C_{L1}C_{R2}\gg C_{R1}C_{L2}$.
A simplified parametrization for $\beta$ assuming $C_{L1}=C_{R2}=C$ and
$C_{L2}=C_{R1}=\eta C$ in Eq.~(\ref{betta}) yields
\beq
\label{param}
\beta=\frac{1-\eta^2}
{(1+\eta)(1+\eta+2\tilde{C}_0)} \;,
\eeq
with $\tilde{C}_0=C_0/C$, see 
Fig.~\ref{Fig:fig4} for a graphical illustration.
A large asymmetry is hence obtained for small values of $\eta \ll 1$,
in addition to a small inter-dot capacitance $C_0$ compared to $C$.
This corresponds effectively to a situation where 
each dot is strongly coupled to
a {\it single} lead, the upper dot to the left 
lead and the lower one to the right,
with negligible couplings to the other lead as 
well as between the two dots \footnote{Notice 
that the orbital Kondo effect requires a minimum 
coupling $\tilde{C}_0$, in order to get a large 
enough effective charging energy $E_c$.}.
Indeed, the results in Fig.~\ref{Fig:fig6} identify
$\lambda_{LL}^{z\ua}$, $\lambda_{LL}^{z\da}$ and $\lambda_{LR}^\perp$ as
leading couplings controlling the physical description in this region of the
phase diagram.
From the Hamiltonian~(\ref{hk1}) a purely orbital 
Kondo effect is therefore expected,
where the $\ua$ (or $\da$) spin configurations binds the upper dot
to the left electrode (the lower dot to the right electrode respectively).
Such device has been analyzed in great detail in Ref.~[\onlinecite{borda}]
and confirms the development of an orbital Kondo effect.

\begin{figure}[t]
   \epsfig{figure=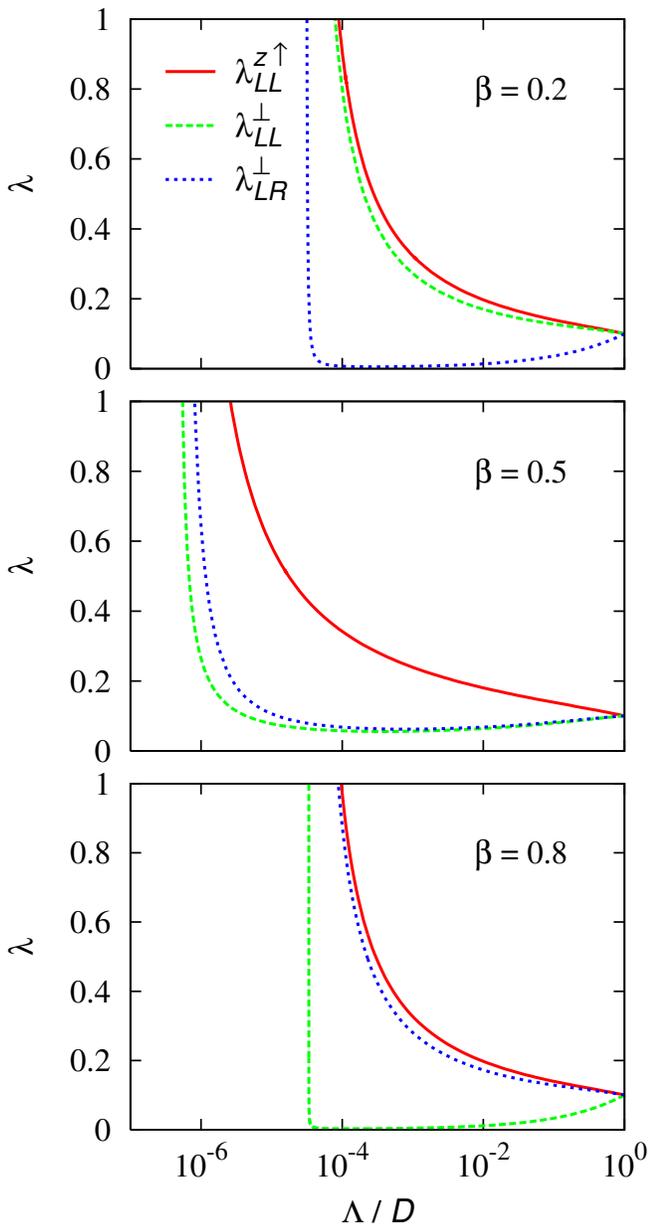,width=8.5cm}
   \caption{(Color online) Flow of the Kondo couplings $\lambda_{LL}^{z\ua}$,
     $\lambda_{LL}^\perp$ and $\lambda_{LR}^\perp$ as a function of $\Lambda$
     for a dissipation $r=1$ and asymmetry parameter $\beta=0.2, 0.5, 0.8$.
     Note the change in the Kondo scale determined by the divergence of
     $\lambda_{LL}^{z\ua}$.}
   \label{Fig:fig7}
\end{figure}

Comparing the behavior of $\lambda_{LL}^\perp$ and $\lambda_{LR}^\perp$
displayed on the third an fifth panel, a symmetry 
under the exchange of $\beta$ in
$1-\beta$ appears, related to the symmetry of the prefactors of $r$ in
Eqs.~(\ref{flow2}c) and (\ref{flow2}e) around $\beta=1/2$.
This symmetry is also visible in the full flow of these couplings as shown
on the three panels of Fig.~\ref{Fig:fig7} 
for $r=1$, associated with small,
intermediate and large $\beta$ respectively.
Note that for smaller values of the dissipation 
$r$ a weak asymmetry arises due to
marginal contributions (in the RG sense) from the flow of the irrelevant
coupling $\lambda_{RL}^\perp$.

The two-channel Kondo phase IV for $\beta<1/2$ 
(at $r>1/2$) is therefore replaced
by an orbital Kondo phase for $\beta>1/2$, and is denoted as III
in Fig.~\ref{Fig:fig3}.
The smooth boundary from II to III is given by 
a vanishing $\lambda_{LL}^\perp$
coupling, described by the condition 
$\beta^2r=1/2$ according to Eq.~(\ref{flow2}c).

\underline{\em Case $\beta\simeq 1/2$.}\\
For $r>2$, all transverse couplings except 
$\lambda_{LL}^{z\ua}$ and $\lambda_{LL}^{z\da}$ 
are driven
irrelevant \footnote{Note that the divergence of 
$\lambda_{LL}^{z\ua}$ in the localized
region is unphysical and is due to the perturbative treatment.}.
Since even now $\lambda_{LL}^\perp$ is driven to $0$, the spin flips, or
equivalently the orbital flips, are completely 
suppressed and the Kondo effect disappears
(formally $T_K=0$).
This qualitatively different phase corresponds to 
a state where the impurity spin is
fully localized in a given orientation by the 
environment. The localized phase of
the model is denoted as the region V in Fig.~\ref{Fig:fig3},
defined by the conditions $\beta^2 r>1/2$ and $(1-\beta)^2 r>1/2$ from
the preceding arguments. These lines correspond to a true vanishing of $T_K$
(from either the III and IV Kondo phases) and define a genuine quantum phase transition.

We infer by analogy with previous works on the 
Kondo effect in a dissipative ohmic environment
that these phase transitions are of 
Kosterlitz-Thouless type.\cite{karyn1,karyn2,BZS}
For $r<2$, we recover the usual spin-orbital Kondo phase (at $r\ll 1$) through
the intermediate crossover region II.

\subsubsection{Kondo temperature $T_K$}

\begin{figure}[ht]
   \epsfig{figure=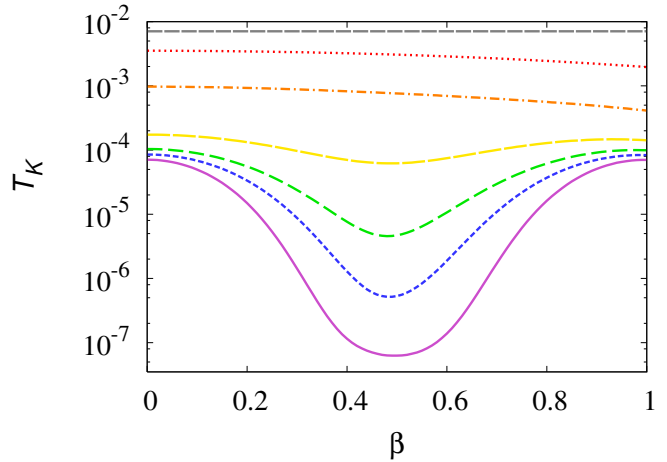,height=8.5cm,angle=-90}
   \caption{(Color online) Kondo temperature $T_K$ 
as a function of asymmetry parameter
     $\beta$, for different values of 
$r=0,0.1,0.25,0.5,0.75,1,1.5$ from top to bottom.}
   \label{Fig:fig8}
\end{figure}

The effect of both environmental dissipation and asymmetry in the
capacitances on the Kondo temperature $T_K$ is 
reported in Fig.~\ref{Fig:fig8}.
The results for the Kondo temperature, defined by the condition
$j_{LL}^{z\ua}(\Lambda\!=\!T_K)=10$, exhibit a 
pronounced non-monotonic behavior
as a function of the asymmetry parameter $\beta$ for increasing values of the
dissipation $r$.
For $\beta\gtrsim 1$ the two-channel Kondo phase 
develops with increasing dissipation,
while for $\beta\lesssim 1$ an orbital Kondo 
phase is observed. For intermediate
values of $\beta$ and for $r>2$, a vanishing 
Kondo temperature indicates the appearance
of the localized phase. In the more realistic case $0<r<1$, the Kondo
temperature is finite, but sharply reduced for $\beta\simeq 1/2$.

\begin{figure}[t]
   \epsfig{figure=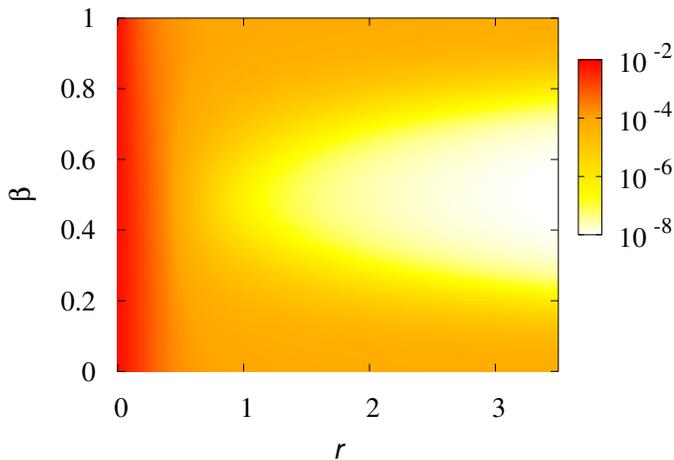,width=6.cm,angle=-90}
   \caption{(Color online) Kondo temperature $T_K$ 
as a function of dissipation $r$
     and asymmetry parameter $\beta$, without 
geometric asymmetry in the tunneling
     amplitudes.}
   \label{Fig:fig9}
\end{figure}

The complete evolution of the Kondo temperature in the
$(r,\beta)$ phase space is provided in Fig.~\ref{Fig:fig9}.

\subsection{Geometric effects: Asymmetric tunneling amplitudes}

We next describe the modifications to the previous phase diagram taking into
account the effect of a geometric asymmetry both on the capacitances
and on the tunneling amplitudes.
In particular, the initial values of the Kondo couplings will be affected
according to Eqs.~(\ref{couplings}).
For the present description in terms of equivalent circuits, we will assume that the
$C_i$'s and $t_i $'s are proportional in determining the initial values of
the couplings.

In this section we will discuss the resulting phase diagram, and the corresponding 
behavior of the Kondo temperature.
Finally, we will address the particular case of strongly asymmetric capacitances
$(\beta \lesssim 1)$ corresponding to two quantum dots in series.

\subsubsection{Phase diagram and Kondo temperature $T_K$}

\begin{figure}[t]
   \epsfig{figure=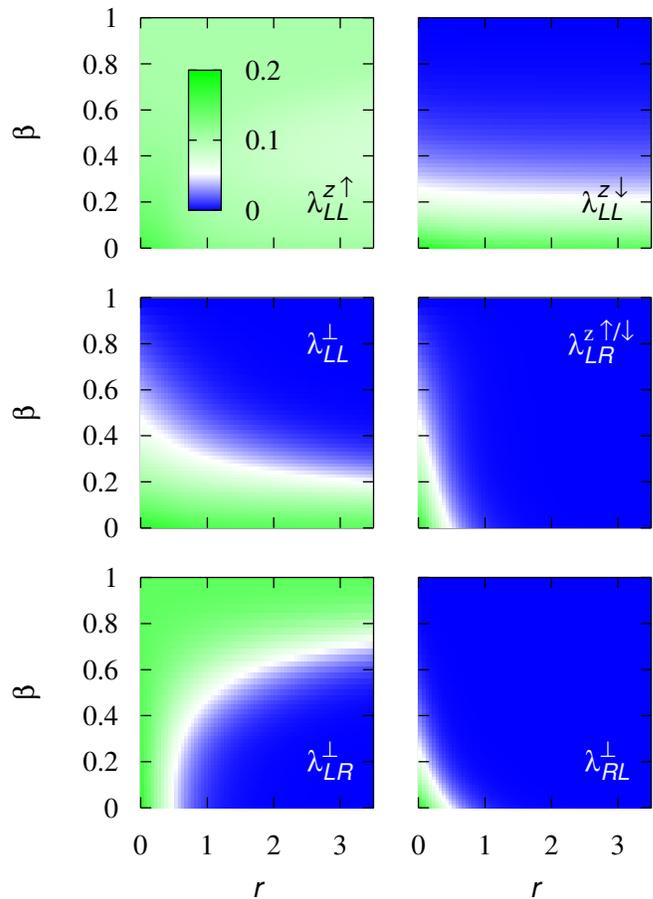,width=8.5cm}
   \caption{(Color online) Typical value of the renormalized Kondo couplings
     $\lambda_i$ according to Eqs.~(\ref{flow2}) for the case of asymmetric
capacitances and tunneling amplitudes (corresponding to asymmetric initial 
conditions for the Kondo couplings).
     These graphs are displayed as a function of dissipation $r$ and asymmetry
     parameter $\beta$ (with parameter $\tilde{C}_0=0.01$). 
The corresponding scale $\Lambda^*$ is determined as in
     Fig.~\ref{Fig:fig6}.}
     \label{Fig:fig10}
\end{figure}

The physical properties can again be determined directly from the behavior of the
different couplings during the flow. The equations describing the flow in
presence of modified tunneling amplitudes due to a capacitance asymmetry are
the same as in the previous treatment, Eqs.~(\ref{flow2}), whereas the
initial conditions are given by Eqs.~(\ref{couplings}).
For the parametrization of the tunneling amplitudes we use the model chosen
above Eq.~(\ref{param}), with $\eta$ and $\tilde{C}_0$ as independent parameters. 
As a consequence, the maximal value of $\beta$ is limited by $\tilde{C}_0$. In order 
to access an extended range for $\beta$, we take $\tilde{C}_0=0.01$ in all
further numerical calculations, larger values lead only to small quantitative modifications. 
Finally, as mentioned above, we assume 
that the geometric asymmetry in the tunneling amplitudes follows the same model
as for the capacitances, namely: $t_{L1}=t_{R2}=t$ and $t_{L2}=t_{R1}=\eta t$.
The initial (bare) Kondo couplings scale according to equation~(\ref{couplings}).
Results for the solution of the flow equations Eqs.~(\ref{flow2}) 
with the above initial conditions
are shown in Fig.~\ref{Fig:fig10} as a function of $\beta$ and $r$.

\begin{figure}[t]
   \epsfig{figure=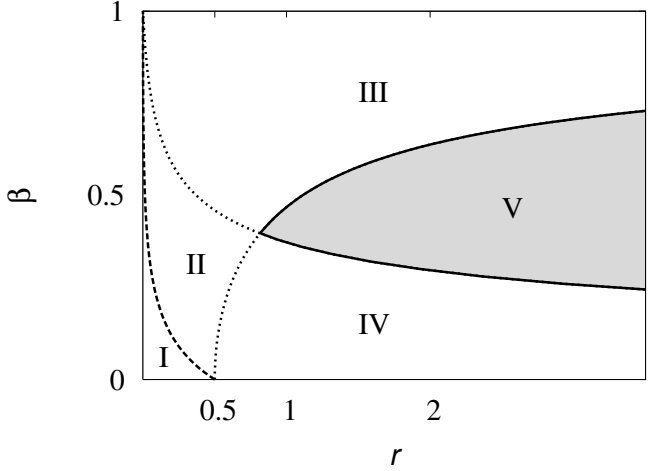,height=8.5cm,angle=-90}
   \caption{Phase diagram as a function of dissipation $r$
     and capacitance asymmetry parameter $\beta$ (assuming $\tilde{C}_0\ll1$), 
     including geometric asymmetry in the tunneling
     amplitudes. The different regions and transition lines are characterized
     by the same physical behavior as for 
     symmetric tunneling amplitudes displayed in
     Fig.~\ref{Fig:fig3}.}
   \label{Fig:fig11}
\end{figure}

\begin{figure}[ht]
   \epsfig{figure=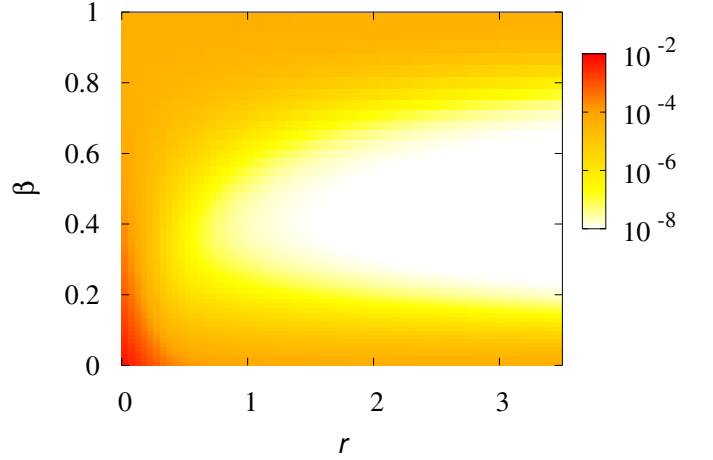,width=6.cm,angle=-90}
   \caption{(Color online) Kondo temperature $T_K$ 
     as a function of dissipation $r$
     and capacitance asymmetry parameter $\beta$, with geometric asymmetry in the tunneling
     amplitudes.}
   \label{Fig:fig12}
\end{figure}

The symmetry $\beta \leftrightarrow 1-\beta$ for large dissipation is not
preserved anymore, in particular a stronger suppression 
of the Kondo couplings occurs in
the region $\beta>1/2$.
This leads to a reduction of region I in favor of regions II and III of the
previous phase diagram for equal initial 
couplings of Fig.~\ref{Fig:fig3}.
However, the topology and qualitative properties are conserved.
An analytical understanding of the flow with asymmetric
initial conditions can be obtained by noting that the initial Kondo
couplings~(\ref{couplings}) are simply dressed by appropriate powers of the
asymmetry parameter $\eta$. A simple rescaling of the Kondo terms by these
$\eta$-dependent prefactors allows to map Eqs.~(\ref{flow2}) with asymmetric couplings onto 
a set of effective flow equations with symmetric initial couplings and modified 
effective scaling dimensions. The previous analysis of the phase diagram allows to 
locate instantaneously the phase transition and crossover lines, that are now 
dressed by geometric asymmetry coefficients. These read:
\begin{subequations}
   \label{new}
   \begin{eqnarray}
     && \eta^{-2}\,(1+\beta)^2r=\frac{1}{4}(1+\eta^2)\\
     &&(1-\beta)^2r=\frac{1}{4}(1+\eta^2) \\
     && \eta^{-1}\,\beta^{\,2}r=\frac{1}{4}(1+\eta^2) \;,
   \end{eqnarray}
\end{subequations}
where Eq.~(\ref{new}a) determines the crossover line between region I and II,
Eq.~(\ref{new}b) between region II and IV, and 
Eq.~(\ref{new}c) between region II
and III. 
In the present case of a weak inter-dot capacitance $\tilde{C}_0 \ll 1$, the
parametrization~(\ref{param}) can be easily inverted as
\beq
\eta=\frac{1-\beta}{1+\beta} 
\eeq
and the use of the relation
\beq
\frac{1}{2} (1+\eta^2)=\frac{1+\beta^2}{(1+\beta)^2}
\eeq
leads to a particularly simple form in terms of $\beta$
for the normalization factor appearing in Eqs.~(\ref{new}).
The corresponding phase diagram is shown in Fig.~\ref{Fig:fig11}
and shows only quantitative differences to the previous phase diagram
Fig.~\ref{Fig:fig3} obtained for symmetric tunneling amplitudes.
The phase diagram Fig.~\ref{Fig:fig11} is confirmed by the behavior of the Kondo 
temperature from the full numerical solution, as shown in Fig.~\ref{Fig:fig12}.

We finally observe that the above phase diagram is expected to be generic 
independently of particle-hole symmetry.

\subsubsection{A special case: serial quantum dots}
\label{sec:serial}

\begin{figure}[b]
   \epsfig{figure=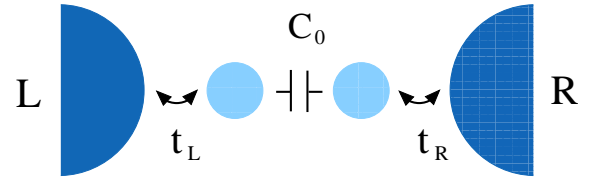,width=7.5cm}
   \caption{(Color online) Schematic representation of a serial DD
     corresponding to the setup proposed in Fig.~\ref{Fig:fig1} in the
     limit of maximal tunneling and capacitance asymmetry.}
   \label{Fig:fig13}
\end{figure}

In this section, we consider the particular case of two quantum dots coupled capacitively 
in series as analyzed in Refs.~[\onlinecite{pohjola,borda,logan,li}].
This corresponds to the limit $\eta = 0$ in the
parametrization of Fig.\ref{Fig:fig4} (i.e. $C_{L2}=C_{R1}=0$) 
and to $t_{L2}=t_{R1}=0$: it leads therefore to a maximal
asymmetry in the capacitances and tunneling amplitudes.
A schematic representation of the resulting serial quantum dot setup is shown
in Fig.~\ref{Fig:fig13}.

The equations describing the flow of the remaining couplings $\lambda_{LL}$
and $\lambda_{LR}$, from the equations for $\lambda_{LL}^{z\ua}$ and
$\lambda_{LR}^{\perp}$ of system~(\ref{flow2}), read
\begin{subequations}
\label{JJdecoh}
\bea
\frac{d \lambda_{LL} }{d \ln \Lambda} &=&-
\frac{1}{2}\left[(\lambda_{LL})^2+(\lambda_{LR})^2\right]\\
\frac{d \lambda_{LR}}{d \ln \Lambda} &=&(1-\beta)^2 r \lambda_{LR}
-\lambda_{LR}\lambda_{LL}\;,
\eea
\end{subequations}
where the capacitance asymmetry parameter reduces 
to $\beta=1/(1+2\tilde{C}_0)$.
The redefinition $\Lambda \to \Lambda^2$ and $2(1-\beta)^2r \to r$ in the above
equations reduces to the system~(\ref{JJ}), that 
describes the flow of the inter-
and intra-lead processes for a single quantum dot in an ohmic environment.
The $(r,\beta)$ phase diagram presents therefore two regimes, separated by the
condition $2(1-\beta)^2r=1/2$. The regions I, II and III of
Fig.~\ref{Fig:fig3} correspond to a single-channel Kondo regime, while regions 
IV and V to a two-channel Kondo regimes.

In addition, the respective Kondo temperature 
appears lowered by $T_K \to (T_K)^2$
with respect to the single dot.

\section{Conclusion and discussion}\label{sec:conclusions}

The present analysis extends the recent 
investigation of Ref.~[\onlinecite{florens06}]
of the effects of electromagnetic noise on the Kondo effect in a quantum dot to
a capacitively coupled DD device, characterized by entangled
spin and orbital degrees of freedom.
The low-energy physics of the above spin-orbital Kondo model in presence of an
ohmic environment of resistance $R$ exhibits a 
rich phase diagram as a function of the
capacitance asymmetry parameter $\beta$ and dimensionless resistance $r=R/R_K$, with
the quantum value $R_K=h/e^2$.
Using a perturbative renormalization-group approach, we could identify several 
crossover regions between different Kondo fixed points and a 
distinctive localized phase (where spin-flip processes are fully suppressed)
for large values of $r$ and $\beta \simeq 1/2$.
The corresponding Kondo temperature displays a pronounced non-monotonic behavior 
with $\beta$, with a strong reduction in proximity of the localized phase.
Additional geometric asymmetries in the tunneling amplitudes further reduce 
spin-flip processes, thus favoring the development of the localized phase.

Finally, the present results lead to interesting consequences in the context of
quantum information theory, in particular on the efficiency of spin filtering
devices in presence of environmental fluctuations.
Promising proposals for a controlled realization of spin qubits and on their
operations rely on the use of entangled spin and orbital degrees of freedom.
In view of an experimental realization, an 
essential issue concerns the spin coherence. The 
present work helps to understand how the phase 
and spin-coherent transmission through the device 
is affected by inelastic interaction with the 
environment. To minimize those effects, the 
device should be as symmetric as possible, in its 
geometry, and operate at low bias voltage.
The combination of recent experimental realizations of both strongly capacitive 
dots\cite{huebel} and a strongly resistive environment\cite{pierre} represents a 
promising prospect.

\acknowledgments{This work has been supported by the contract
ANR\_05\_NANO\_050\_S2.
}

\appendix

\section{Complete set of RG equations}\label{sec:appendix}

The RG flow equations of the ten Kondo couplings 
are derived in straightforward manner
paying attention to the fact that most of them 
acquire some anomalous dimensions.
At second order in $\lambda$, the RG flow reads
\begin{widetext}
\begin{subequations}
\label{flow1}
\bea
\frac{d \lambda_{LL}^{z\ua} }{d \ln \lambda} &=&-
\frac{1}{2}\left[(\lambda_{LL}^{z\ua})^2+(\lambda_{LL}^\perp)^2
+(\lambda_{LR}^{z\ua})^2+(\lambda_{LR}^\perp)^2\right]
\\
\frac{d \lambda_{LL}^{z\da}}{d \ln \lambda} &=&-
\frac{1}{2}\left[(\lambda_{LL}^{z\da})^2+(\lambda_{LL}^\perp)^2
+(\lambda_{LR}^{z\da})^2+(\lambda_{RL}^\perp)^2\right]
\\
\label{asym} \frac{d \lambda_{LL}^\perp}{d \ln 
\lambda} &=&(\kappa_{L2}-\kappa_{L1})^2 r
\lambda_{LL}^\perp-\frac{1}{2}\left[\lambda_{LL}^\perp(\lambda_{LL}^{z\ua}+\lambda_{LL}^{z\da})
+\lambda_{LR}^\perp\lambda_{LR}^{z\da}+\lambda_{RL}^\perp\lambda_{LR}^{z\ua}\right]
\\
\frac{d \lambda_{LR}^{z\ua}}{d \ln \lambda} 
&=&(\kappa_{L1}+\kappa_{R1})^2 r 
\lambda_{LR}^{z\ua}
-\frac{1}{2}\left[\lambda_{LR}^{z\ua}(\lambda_{LL}^{z\ua}+\lambda_{RR}^{z\ua})
+\lambda_{LL}^\perp\lambda_{RL}^\perp 
+\lambda_{RR}^\perp\lambda_{LR}^\perp\right]
\\
\frac{d \lambda_{LR}^{z\da}}{d \ln \lambda} 
&=&(\kappa_{L2}+\kappa_{R2})^2 r 
\lambda_{LR}^{z\da}
-\frac{1}{2}\left[\lambda_{LR}^{z\da}(\lambda_{LL}^{z\da}+\lambda_{RR}^{z\da})
+\lambda_{LL}^\perp\lambda_{LR}^\perp 
+\lambda_{RR}^\perp\lambda_{RL}^\perp\right]
\\
\frac{d \lambda_{LR}^{\perp}}{d \ln \lambda} 
&=&(\kappa_{L1}+\kappa_{R2})^2 r 
\lambda_{LR}^{\perp}
-\frac{1}{2}\left[\lambda_{LR}^{\perp}(\lambda_{LL}^{z\ua}+\lambda_{RR}^{z\da})+
\lambda_{LL}^{\perp}\lambda_{LR}^{z\da}+\lambda_{RR}^{\perp}\lambda_{LR}^{z\ua}\right]
\\
\frac{d \lambda_{RL}^{\perp}}{d \ln \lambda} 
&=&(\kappa_{R1}+\kappa_{L2})^2 r 
\lambda_{RL}^{\perp}
-\frac{1}{2}\left[\lambda_{RL}^{\perp}(\lambda_{LL}^{z\da}+\lambda_{RR}^{z\ua})+
\lambda_{LL}^{\perp}\lambda_{LR}^{z\ua}+\lambda_{RR}^{\perp}\lambda_{LR}^{z\da}\right]\;,
\eea
\end{subequations}
\end{widetext}
where the parameter $r=R/R_K$ defines the normalized resistance.
The RG equations for the Kondo couplings 
$\lambda_{RR}^{z\ua/\da,\perp}$ are simply
inferred from $\lambda_{LL}^{z\ua/\da,\perp}$ by 
exchanging $L\leftrightarrow R$.
Though $\lam_{\ga\ga}^{z\ua}$ may be equal to 
$\lam_{\ga\ga}^{z\da}$ at the bare level,
some asymmetry may be generated through the renormalization of the couplings
$J_{LR}^{z\ua/\da}$.
Similarly, a finite prefactor 
$(\kappa_{L2}-\kappa_{L1})^2$ of $r$ in 
Eq.~(\ref{asym})
can arise dynamically in the flow even for equal initial conditions, leading to
a suppression of the orthogonal coupling 
$\lambda_{LL}^\perp$. The disappearance of the
Kondo effect in this situation leads to a 
qualitatively different picture as in the
noisy single-dot setup discussed previously.\cite{florens06}

An important question concerns the effect of the frequency dependence of the
circuit's impedance, neglected in the present treatment.
The full energy distribution function $P(E)$ of the environmental modes
for an ohmic dissipation can be taken into account within the $P(E)$ theory.
For a single quantum dot of Sec.~\ref{sec:sym}, 
the main modification of the flow
equations~(\ref{JJ}) consists in the
vanishing of the dissipative term appearing in Eq.~(\ref{JB}) above an energy
cutoff $\omega_c\simeq 1/(RC)$, where $R$ is the resistance of the environment
and $C$ the effective capacitance of the equivalent circuit.
The absence of DCB for $\Lambda>\omega_c$ leads to an initial renormalization
of all couplings without environmental effects; for $\Lambda<\omega_c$ the
dissipative low-energy behavior comes into play.
Concerning the associated Kondo temperature, in 
addition to the dissipation effects
a sensitive dependence on the full spectral 
function of excited environmental modes is
observed. In particular,
the onset of DCB at scale $\w_c$ implies in general a less pronounced decrease
of the associated Kondo temperature for 
increasing dissipation than previously predicted.
For a more detailed analysis we refer to Ref.~[\onlinecite{florens06}].


\begin{thebibliography}{99}

% QBITS
\bibitem{loss98} D. Loss and D.P. DiVincenzo, 
Phys. Rev. A {\bf 57}, 120 (1998).
\bibitem{expt-spinqubits} E.A. Laird,  J.R. 
Petta, A.C. Johnson, C.M. Marcus, A. Yacoby, M.P. 
Hanson, and A.C. Gossard, Phys. Rev. Lett. {\bf 
97}, 056801 (2006);
F.H.L. Koppens, C. Buizert, K.J. Tielrooij, I.T. 
Vink, K.C. Nowack, T. Meunier, L.P. Kouwenhoven, 
and L.M.K. Vandersypen, Nature {\bf 442}, 766 
(2006).
\bibitem{entangled_pairs} P. Recher,  E.V.
Sukhorukov, and D. Loss, Phys. Rev. B {\bf 63},
165314 (2001); O. Sauret, T. Martin, and D.
Feinberg, Phys. Rev. B {\bf 72}, 024544 (2005).
\bibitem{teleportation} O. Sauret, D. Feinberg,
and T. Martin, Phys. Rev. B {\bf 69}, 035332 (2004).
\bibitem{bychkov} A. Bychkov and E.I. Rashba, J. Phys. C {\bf 17} (2002).
\bibitem{nitta02} T. Koga, J. Nitta, T. Akazaki, 
and H. Takayanagi, Phys. Rev. Lett. {\bf 89}, 
046801 (2002).
\bibitem{precession}
T. Koga, J. Nitta, and M. van Veenhuizen,
Phys. Rev. B {\bf 70}, 161302 (2004);
T. Koga, Y. Sekine, and J. Nitta, Phys. Rev. B {\bf 74}, 041302(R) (2006).
\bibitem{apl} D. Feinberg and P. Simon, Appl. 
Phys. Lett. {\bf 85}, 1846  (2004).
\bibitem{sg} P. Simon and D. Feinberg, Phys. Rev. 
Lett. {\bf 97}, 247207 (2006).

% REVIEWS DYNAMICAL CB
\bibitem{IN} G.-L. Ingold and Yu. V. Nazarov, in "Single Charge
Tunneling", edited by H. Grabert and M. H. Devoret, NATO ASI Series B,
Vol. 294, pp. 21-107 (Plenum Press, New York, 1992).
\bibitem{devgrab} M.~H. Devoret and H. Grabert, in "Single Charge
Tunneling", edited by H. Grabert and M. H. Devoret, NATO ASI Series B,
Vol. 294, pp. 1-19 (Plenum Press, New York, 1992).
% Decoherence DD
\bibitem{aguado} R. Aguado and 
L.P. Kouwenhoven, Phys. Rev. Lett. {\bf 84}, 1986 
(2000).\bibitem{bruder} F. Marquardt and C. 
Bruder, Phys. Rev. B {\bf 68}, 195305 
(2003).\bibitem{dupont} E. Dupont and K. Le Hur, 
Phys. Rev. B {\bf 73} 045325 (2006).
% Kondo+ Noise
\bibitem{karyn1} K. Le Hur, Phys. Rev. Lett. {\bf 92}, 196804 (2004).
\bibitem{karyn2} M.-R. Li, K. Le Hur, and W. 
Hofstetter, Phys. Rev. Lett. {\bf 95}, 086406 
(2005).
\bibitem{BZS} L. Borda, G. Zarand, and P. Simon, Phys. Rev. B {\bf 72}, 155311
(2005).
\bibitem{BZGG} L. Borda, G. Zarand, and D. Goldhaber-Gordon, cond-mat/0602019.
\bibitem{kaminski00} A. Kaminski, Yu.V. Nazarov, and L.I. Glazman,
   Phys. Rev. B {\bf 62}, 8154 (2000).
\bibitem{lopez01}  R. L\'opez, R. Aguado, G. 
Platero, and C. Tejedor, Phys. Rev. B {\bf 64}, 
075319 (2001).
\bibitem{florens06} S. Florens, P. Simon, S. 
Andergassen, and D. Feinberg, Phys.
Rev. B {\bf 75}, 155321 (2007).

%Capacitively coupled quantum dots
\bibitem{borda} L. Borda, G. Zarand, W. 
Hofstetter, B.I. Halperin, and J. von Delft, 
Phys. Rev. Lett. {\bf 90}, 026602  (2003).
\bibitem{pohjola} T. Pohjola, H. Schoeller, and 
G. Sch\"on, Europhys.  Lett. {\bf54}, 241 (2001).
\bibitem{logan} M.R. Galpin, D.E. Logan, and H.R. Krishnamurthy,
Phys. Rev. Lett. {\bf 94}, 186406 (2005); {\it 
ibid}, J. Phys.: Condens. Matter {\bf 18}, 6545 
(2006).
\bibitem{li} M.-R. Li and K. Le Hur,
Phys. Rev. Lett. {\bf 93}, 176802 (2004).

\bibitem{recher} P. Recher, E.V. Sukhorukov, and D.  Loss, Phys.
Rev. Lett. {\bf 85}, 1962 (2000).
\bibitem{hanson} R. Hanson, L.H.Willems vanBeveren, I.T. Vink, J.M. Elzerman, W.J.M. Naber, 
F.H.L. Koppens, L.P. Kouwenhoven, and L.M.K. 
Vandersypen, Phys. Rev. Lett. {\bf 94}, 196802 
(2005).
% Some SU(4)
\bibitem{lehur03} K. Le Hur and P. Simon, Phys. 
Rev. B {\bf 67}, 201308(R) (2003); K. Le Hur, P. 
Simon, and L. Borda, Phys. Rev. B {\bf 69}, 
045326 (2004);  K. Le Hur, P. Simon, and D. Loss, 
Phys. Rev. B {\bf 75}, 035332 (2007).
\bibitem{jarillo} P. Jarillo-Herrero, J. Kong, H. 
S. J. van der Zant, C. Dekker, L. P. Kouwenhoven, 
and S. De Franceschi, Nature {\bf 434}, 484 
(2005).
\bibitem{choi05} M.-S. Choi, R. Lopez, and R. 
Aguado, Phys. Rev. Lett. {\bf 95}, 067204 (2005).

\bibitem{martinek} J. Martinek, Y. Utsumi, H. 
Imamura, J. Barnas, S. Maekawa, J. K\"onig, and 
G. Sch\"on, Phys. Rev. Lett. {\bf 91},  127203 
(2003); M.-S. Choi, D. Sanchez and R. Lopez, 
Phys. Rev. Lett.  {\bf 92}, 056601 (2004); P. 
Simon, P.S. Cornaglia, D. Feinberg, and C.A. 
Balseiro, Phys. Rev. B {\bf 75}, 045310 (2007).

% Papers on DCB
\bibitem{grabert91} H. Grabert, G.-L. Ingold, 
M.H. Devoret, D. Est\'eve, H. Pothier, and C. 
Urbina, Z. Phys. B {\bf 84}, 143 (1991); G.-L. 
Ingold, P. Wyrowski, and H. Grabert, Z. Phys. B 
{\bf 85}, 443  (1991).
\bibitem{odintsov} A. A. Odintsov, V. Bubanja, 
and G. Sch\"on, Phys. Rev. B {\bf 46}, 6875 
(1992).
\bibitem{nazarov91} U. Geigenm\"uller and Yu.V. 
Nazarov, Phys. Rev. B {\bf 44}, 10953 (1991).

% Experiments
\bibitem{huebel} A. H\"ubel, J. Weiss, W. Dietsche, and K. v. Klitzing,
App. Phys. Lett. {\bf 91}, 102101 (2007).
\bibitem{pierre} C. Altimiras, U. Gennser, A. Cavanna, D. Mailly, and F. Pierre, to be published in 
Phys.Rev. Lett. (2007).

\end{thebibliography}
\end{document}